# Markov Process Simulation with Quantum Computer

Petar N. Nikolov

FDIBA, Technical University - Sofia
Sofia, Bulgaria, petar.nikolov@fdiba.tu-sofia.bg

*Abstract* — This paper shows a novel way of simulating a Markov process by a quantum computer. The main purpose of the paper is to show a particular application of quantum computing in the field of stochastic processes analysis. Using a Quantum computer, the process could be superposed, where the random variables of the Markov chain are represented by entangled qubit states, which gives the great opportunity of having all the possible scenarios simultaneously.

## I. Markov Process Basics

Here some of the fundamentals, standing behind the theory for a Markov process, are stated. A stochastic process is a collection of random variables from some probability space into a state space [1]. By definition a Markov process is a random process, whose future probabilities are determined by its most recent state [2,3]. A stochastic process **x(t)** is called Markov if for every **n** and $t_1<t_2<...<t_n$ :

$$P(x(tn) \le x_n | x(t_{(n-1)}),...,x(t_1)) = P(x(t_n) \le x_n | x(t_{(n-1)})). \quad (1)$$

The Markovian process realization in a given period only depends on the previous period realization (1).

There are two objects, characterizing the Markovian property of a discrete stochastic process **{X}** with **n** possible states:

- The transition matrix Π. This matrix describes the transition from one position in the state space to another with every move's probability. The size of the matrix is $n \times n$.

$$\Pi_{(i,j)} = P(X_{(j,t-1)} | X_{(i,t)}) \quad (2)$$

where $\Pi_{(i,j)}$ is an element of the transition matrix $\Pi$, which gives the probability of moving from position **i** into position **j**.

- Vector π$_t$, which describes the probability to be at every possible position in the state space at time **t** (3). The size of the vector is $1 \times n$.

So for any $t>0$,

$$\pi_{(t,i)} \ge 0 \text{ for } \forall t,i \text{ and } \sum_{i=1}^{n} \pi_{(t,i)} = 1 \quad (3)$$

Since Markov process is a stochastic process which possesses Markov property, irrespective of the nature of the time and state space parameter (discrete or continuous), there are four categories of Markov processes:

- continuous time parameter, continuous state space - continuous parameter Markov chain (**CTMC**)
- continuous time parameter, discrete state space
- discrete time parameter, continuous state space
- discrete time parameter, discrete state space - discrete parameter Markov chain (**DTMC**)

The simplest model of a Markov chain is in the case, where the random variables are pairwise independent, which makes this model rather restrictive.

For the purpose of this paper only Markov chains, which consist of only three random variables, have been reviewed.

## II. Quantum Computing Basics

The quantum bits (qubits) are represented in a similar to the classical bit way. Their value could be either 0 or 1 when measured (the collapse to classical state). However being in a quantum state, a quantum bit is in a superposition of the two states 0 and 1.

$$|\psi> = a_0|0> + a_1|1> \quad (4)$$

This superposition is a linear combination of two orthonormal basis vectors $|0>$ and $|1>$ in two-dimensional state space in C2 (4). Where $a_0$ and $a_1$ are the quantum probabilities (complex numbers), which represent the chance that a give quantum state will be observed when the superposition has collapsed. They satisfy the conditions:

$$|a_0^2| + |a_1^2| = 1 \quad (5)$$

$$||\psi>|| = <\psi|\psi> = 1 \quad (6)$$

A superposition of a qubit could be represented also by:

$$|s> = \cos(\frac{1}{2}\Theta)|1> + e^{i\varphi}\sin(\frac{1}{2}\Theta)|0> \quad (7)$$

Connecting the qubits together gives a quantum register as a result, where the length of the string determines the amount of the information that could be stored in that register. The register also could be in a superposition (8), which means that all the qubits that construct it are superposed simultaneously.

$$|\psi_n> = \sum_{i=0}^{n} a_i |i> \quad (8)$$

An **n**-bit is in superposition of all the $2^n$ possible bit strings [4,5] (8).

Performing an operation over a quantum register requires the use of a quantum logic gates. Quantum logic gates applied to a quantum register maps one quantum superposition to another, allowing the evolution of the system's state. The operations mathematically represented are tensor product of the transformation matrix of the quantum logic gate with the matrix representation of the register [6].

In this paper mainly two quantum logic gates have been used, so the focus would be on them:

1. $\sqrt[n]{X}$ (**n-root of X-gate**)

The **n-root of X-gate** can be easily constructed by knowing the matrix representations of:

- the Hadamard

$$H = \frac{1}{\sqrt{2}} \begin{bmatrix} 1 & 1 \\ 1 & -1 \end{bmatrix} \qquad (9)$$

- the X-gate (NOT)

$$X = \begin{bmatrix} 0 & 1 \\ 1 & 0 \end{bmatrix} \qquad (10)$$

- **n-root of Z gate**, also known as **Phase shift gate ($R_\varphi$)**

$$\sqrt[n]{Z} = \begin{bmatrix} 1 & 0 \\ 0 & e^{\frac{i\pi}{n}} \end{bmatrix} \qquad (11)$$

and following the rule: Bracketing a Z-axis rotation with Hadamard gates transforms the Z rotation into X rotation [5]:

$$H \sqrt[n]{Z} H = \sqrt[n]{X} \qquad (12)$$

Some mathematics showing the matrix representation of **n-root of X-gate** according to the statements above:

$$\sqrt[n]{X} = H \sqrt[n]{Z} H = \frac{1}{\sqrt{2}} \begin{bmatrix} 1 & 1 \\ 1 & -1 \end{bmatrix} \begin{bmatrix} 1 & 0 \\ 0 & e^{\frac{i\pi}{n}} \end{bmatrix} \frac{1}{\sqrt{2}} \begin{bmatrix} 1 & 1 \\ 1 & -1 \end{bmatrix} =$$

$$= \frac{1}{\sqrt{2}} \frac{1}{\sqrt{2}} \begin{bmatrix} 1 & e^{\frac{i\pi}{n}} \\ 1 & -e^{\frac{i\pi}{n}} \end{bmatrix} \begin{bmatrix} 1 & 1 \\ 1 & -1 \end{bmatrix} = \frac{1}{2} \begin{bmatrix} 1+e^{\frac{i\pi}{n}} & 1-e^{\frac{i\pi}{n}} \\ 1-e^{\frac{i\pi}{n}} & 1+e^{\frac{i\pi}{n}} \end{bmatrix} \qquad (13)$$

$$\sqrt[n]{X} = \frac{1}{2} \begin{bmatrix} 1+e^{\frac{i\pi}{n}} & 1-e^{\frac{i\pi}{n}} \\ 1-e^{\frac{i\pi}{n}} & 1+e^{\frac{i\pi}{n}} \end{bmatrix} \qquad (14)$$

The quantum probabilities for this gate should be as follows, depending on the initial state of the qubit (respectively $|0\rangle$ or $|1\rangle$):

$$|a_0^2| = \left|\frac{1+e^{\frac{i\pi}{n}}}{2}\right|^2 \qquad |a_0^2| = \left|\frac{1-e^{\frac{i\pi}{n}}}{2}\right|^2$$

$$|a_1^2| = \left|\frac{1-e^{\frac{i\pi}{n}}}{2}\right|^2 \qquad |a_1^2| = \left|\frac{1+e^{\frac{i\pi}{n}}}{2}\right|^2 \qquad (15)$$

Just for example, when n=1 or n=2, this is actually the Pauli-X or Square root of X gate respectively:

$$X = \begin{bmatrix} 0 & 1 \\ 1 & 0 \end{bmatrix} \qquad (16)$$

$$\sqrt[2]{X} = \frac{1}{2} \begin{bmatrix} 1+e^{\frac{i\pi}{2}} & 1-e^{\frac{i\pi}{2}} \\ 1-e^{\frac{i\pi}{2}} & 1+e^{\frac{i\pi}{2}} \end{bmatrix} = \begin{bmatrix} \frac{1+i}{2} & \frac{1-i}{2} \\ \frac{1-i}{2} & \frac{1+i}{2} \end{bmatrix} \qquad (17)$$

2. $C\sqrt[n]{X}$ (**Controlled n-root of X-gate**)

The $C\sqrt[n]{X}$ could be represented like any controlled-U gate, where **U** is **n-root of X-gate**. It operates on two qubits, where first qubit serves the role of a control and it has the following matrix representation:

$$C\sqrt[n]{X} = \begin{bmatrix} 1 & 0 & 0 & 0 \\ 0 & 1 & 0 & 0 \\ 0 & 0 & \frac{1+e^{\frac{i\pi}{n}}}{2} & \frac{1-e^{\frac{i\pi}{n}}}{2} \\ 0 & 0 & \frac{1-e^{\frac{i\pi}{n}}}{2} & \frac{1+e^{\frac{i\pi}{n}}}{2} \end{bmatrix} \qquad (18)$$

III. THE APPLICATION OF QUANTUM COMPUTING

Lets have the three random variables **X**,**Y**,**Z**, so that their joint distribution $P_{XYZ}$ forms a Markov chain with order:
**X** → **Y** → **Z** .

All the information about **Z** is contained in **Y**, so there is no need to remember **X**, if the task is to determine **Z**. Every variable has two possible states – **0** and **1**. And there are preliminary conditions, which set each of them being in superposition of the two possible states. They are dependent on each other in the following way: the state of **X** changes the probability of **Y** being in one of the two states, and the state of **Y** changes the probability of **Z**.

Since the chain is constructed only by three random variables, and the possible movement is only going forward, after three steps it will not be possible to move forward anymore. It is interesting to know what is the distribution of the possible paths for going from **X** to **Z**. All the possible paths are $2^3$ → **8 in total** (every variable has only two possible states):
|000>, |001>, |010>, |011>, |100>, |101>, |110>, |111>. If the probabilities that a random variable is in state **0** or **1** are respectively α and β, six different states could be distinguished – **X_α , X_β ; Y_α , Y_β ; Z_α , Z_β** .

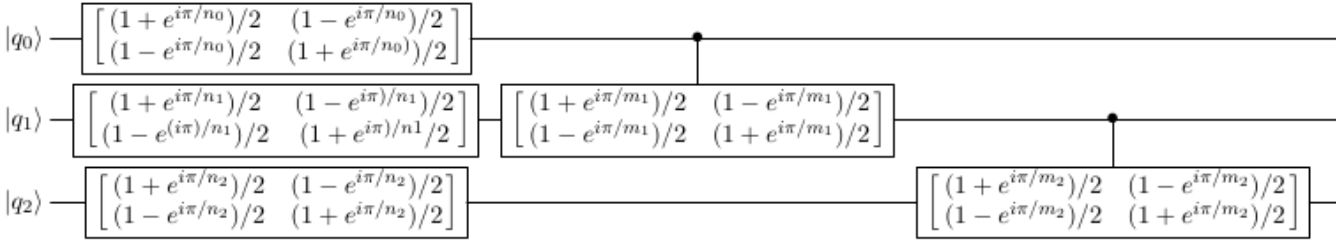

Fig. 1 Quantum circuit, which represents a Markov process

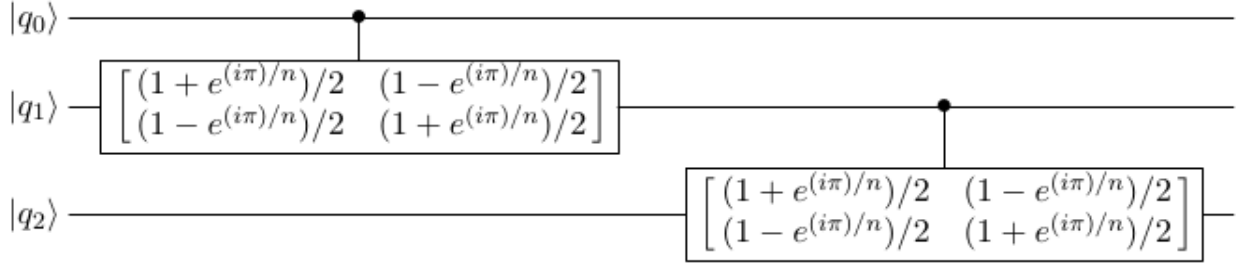

Fig. 2 Quantum circuit, which represents a Markov process without preliminary conditions (the initial state of at least one of the qubits $q_0$ and $q_1$ should be |1>, otherwise none of the controlled gates would fire)

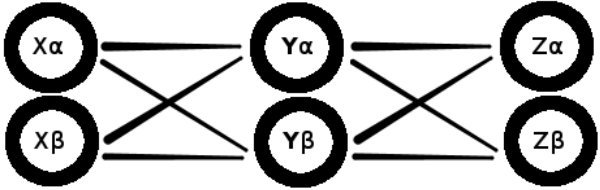

Fig. 3 Schematic of the possible paths for the Markov chain

The quantum circuit shown on Fig. 1 represents the Markov chain model. It is constructed by a quantum register with three qubits – q0, q1, q2, which represent the three random variables X,Y and Z respectively. The state input of the circuit is the computational basis state, consisting of all |0>s . For the purpose of this paper, the quantum circuit could be divided by three stages:

1. The first stage consists the input ( computational basis state in the case of Fig. 1) and the $\sqrt[n_i]{X}$ (**$n_i$-root of X-gate**) - this gate is applied on every qubit and it accomplishes a simple, but useful task – it assign the probabilities, representing the chance that a given quantum state will be observed when the superposition has collapsed (if the circuit has no continuation ).

2. The second stage represents the dependency of the **Y** variable on **X**. $C\sqrt[m_1]{X}$ (**Controlled $m_1$-root of X-gate**) gate acts on the $q_1$ when the control qubit ( $q_0$) is in state |1>. When the control fires, the quantum probabilities $|a_0^2|$ and $|a_1^2|$ for the qubit $q_1$ change to some desired distribution, while still comply with the condition (5).

3. The second stage represents the dependency of the **Z** variable on **Y**. $C\sqrt[m_2]{X}$ (**Controlled $m_2$-root of X-gate**) gate acts on the $q_2$ when the control qubit ( $q_1$) is in state |1>. When the control fires, the quantum probabilities $|a_0^2|$ and $|a_1^2|$ for the qubit $q_2$ change to some desired distribution, while still comply with the condition in (5).

On the Fig. 3 the schematic of all the possible transitions between the states is shown, where the circles represent the possible states, and the lines represent the possible transitions. If:
**$X_α$** = 0 ; **$X_β$** = 1 ; **$Y_α$** = 0 ; **$Y_β$** = 1 ; **$Z_α$** = 0 ; **$Z_β$** = 1,
then the encoding for the quantum superpositions in this case would look like:

**$X_α$** = 0 ; **$X_β$** = 1 ; **$Y_α$** = 0 ; **$Y_β$** = 1 ; **$Z_α$** = 0 ; **$Z_β$** = 1
- **$X_α$ → $Y_α$ → $Z_α$** = |**000**>
- **$X_α$ → $Y_α$ → $Z_β$** = |**001**>
- **$X_α$ → $Y_β$ → $Z_α$** = |**010**>
- **$X_α$ → $Y_β$ → $Z_β$** = |**011**>
- **$X_β$ → $Y_α$ → $Z_α$** = |**100**>
- **$X_β$ → $Y_α$ → $Z_β$** = |**101**>
- **$X_β$ → $Y_β$ → $Z_α$** = |**110**>
- **$X_β$ → $Y_β$ → $Z_β$** = |**111**>

To describe the system with the three qubits equation (8) is used:

$$|\psi_3> = a_0|000> + a_1|001> + a_2|010> + a_3|011> + \\ + a_4|100> + a_5|101> + a_6|110> + a_7|111> \quad (19)$$

where $a_i$ is the amplitude of the state |i> .

As mentioned before, the first step is to initialize the system to 0:

$$1|000> \quad (20)$$

and then apply the $\sqrt[n_i]{X}$ to every qubit to obtain the amplitudes associated with each state:

$$\sqrt[n_i]{\overline{X}}^3|000>=\left(\frac{1+e^{\frac{i\pi}{n_0}}}{2}\right)\left(\frac{1+e^{\frac{i\pi}{n_1}}}{2}\right)\left(\frac{1+e^{\frac{i\pi}{n_2}}}{2}\right)|000>+$$

$$+\left(\frac{1+e^{\frac{i\pi}{n_0}}}{2}\right)\left(\frac{1+e^{\frac{i\pi}{n_1}}}{2}\right)\left(\frac{1-e^{\frac{i\pi}{n_2}}}{2}\right)|001>+$$

$$+\left(\frac{1+e^{\frac{i\pi}{n_0}}}{2}\right)\left(\frac{1-e^{\frac{i\pi}{n_1}}}{2}\right)\left(\frac{1+e^{\frac{i\pi}{n_2}}}{2}\right)|010>+$$

$$+\left(\frac{1+e^{\frac{i\pi}{n_0}}}{2}\right)\left(\frac{1-e^{\frac{i\pi}{n_1}}}{2}\right)\left(\frac{1-e^{\frac{i\pi}{n_2}}}{2}\right)|011>+$$

$$+\left(\frac{1-e^{\frac{i\pi}{n_0}}}{2}\right)\left(\frac{1+e^{\frac{i\pi}{n_1}}}{2}\right)\left(\frac{1+e^{\frac{i\pi}{n_2}}}{2}\right)|100>+$$

$$+\left(\frac{1-e^{\frac{i\pi}{n_0}}}{2}\right)\left(\frac{1+e^{\frac{i\pi}{n_1}}}{2}\right)\left(\frac{1-e^{\frac{i\pi}{n_2}}}{2}\right)|101>+$$

$$+\left(\frac{1-e^{\frac{i\pi}{n_0}}}{2}\right)\left(\frac{1-e^{\frac{i\pi}{n_1}}}{2}\right)\left(\frac{1+e^{\frac{i\pi}{n_2}}}{2}\right)|110>+$$

$$+\left(\frac{1-e^{\frac{i\pi}{n_0}}}{2}\right)\left(\frac{1-e^{\frac{i\pi}{n_1}}}{2}\right)\left(\frac{1-e^{\frac{i\pi}{n_2}}}{2}\right)|111>=|\psi>$$

(21)

The next step combines stage two and three, described in the previous paragraph – it applies the $C\sqrt[m]{\overline{X}}$ to the qubits $q_1$ and $q_2$.

$$|\psi>=\left(\frac{1+e^{\frac{i\pi}{n_0}}}{2}\right)\left(\frac{1+e^{\frac{i\pi}{n_1}}}{2}\right)\left(\frac{1+e^{\frac{i\pi}{n_2}}}{2}\right)|000>+$$

$$+\left(\frac{1+e^{\frac{i\pi}{n_0}}}{2}\right)\left(\frac{1+e^{\frac{i\pi}{n_1}}}{2}\right)\left(\frac{1-e^{\frac{i\pi}{n_2}}}{2}\right)|001>+$$

$$+\left(\frac{1+e^{\frac{i\pi}{n_0}}}{2}\right)\left(\frac{1-e^{\frac{i\pi}{n_1}}}{2}\right)\left(\frac{1+e^{\frac{i\pi}{n_2}+\frac{i\pi}{m_2}}}{2}\right)|010>+$$

$$+\left(\frac{1+e^{\frac{i\pi}{n_0}}}{2}\right)\left(\frac{1-e^{\frac{i\pi}{n_1}}}{2}\right)\left(\frac{1-e^{\frac{i\pi}{n_2}+\frac{i\pi}{m_2}}}{2}\right)|011>+$$

$$+\left(\frac{1-e^{\frac{i\pi}{n_0}}}{2}\right)\left(\frac{1+e^{\frac{i\pi}{n_1}+\frac{i\pi}{m_1}}}{2}\right)\left(\frac{1+e^{\frac{i\pi}{n_2}}}{2}\right)|100>+$$

$$+\left(\frac{1-e^{\frac{i\pi}{n_0}}}{2}\right)\left(\frac{1+e^{\frac{i\pi}{n_1}+\frac{i\pi}{m_1}}}{2}\right)\left(\frac{1-e^{\frac{i\pi}{n_2}}}{2}\right)|101>+$$

$$+\left(\frac{1-e^{\frac{i\pi}{n_0}}}{2}\right)\left(\frac{1-e^{\frac{i\pi}{n_1}+\frac{i\pi}{m_1}}}{2}\right)\left(\frac{1+e^{\frac{i\pi}{n_2}}}{2}\right)|110>+$$

$$+\left(\frac{1-e^{\frac{i\pi}{n_0}}}{2}\right)\left(\frac{1-e^{\frac{i\pi}{n_1}+\frac{i\pi}{m_1}}}{2}\right)\left(\frac{1-e^{\frac{i\pi}{n_2}+\frac{i\pi}{m_2}}}{2}\right)|111>$$

(22)

As a result, a complex quantum entangled state, which describes the Markov system, and gives the distributed probabilities for every possible path through the chain, has been achieved.

## CONCLUSION

This novel method based on quantum computing theory, which finds all the possible solutions for problems, being described with Markov models, reveals interesting interdisciplinary connections. Describing the events ( random variables) from the Markov chain with qubits, gives the ability to create an entangled quantum state, which opens a new feasible way of solving problems with complex event dependencies.

## ACKNOWLEDGMENT

The author acknowledge the Institute for Information Management in Engineering ( IMI) for their support during his exchange program at Karlsruhe Institute of Technology ( KIT) and the access to all the necessary information sources for this paper. All the views expressed are those of the author and do not reflect the official position or policy of KIT or IMI.

Some of the figures in this paper were produced with the QASM Circuit viewer: *qasm2circ v1.4.*